\documentclass[a4paper,10pt,twoside]{cpc-hepnp}
\usepackage{CJK,upgreek,fancyhdr}
\usepackage{multicol}
\usepackage{graphicx}
\usepackage{booktabs}
\usepackage{amssymb,bm,mathrsfs,bbm,amscd}
\usepackage[tbtags]{amsmath}
\usepackage{lastpage}
\usepackage{multirow}
\usepackage{color}

\begin{document}
\begin{CJK*}{GB}{gbsn}

\fancyfoot[C]{\small 010201-\thepage}

\footnotetext[0]{Received xxxx 201x}

\title{Recent progress in multiple chiral doublet bands
\thanks{This work was supported by the Shandong Natural Science Foundation, China (Grant No. JQ201701), and the National Natural Science Foundation of China (Grant No. 11622540).}}

\author{%
Shou-Yu Wang\email{sywang@sdu.edu.cn}%
}
\maketitle

\address{
Shandong Provincial Key Laboratory of Optical Astronomy and Solar-Terrestrial
Environment, School of Space Science and Physics,\\ Institute of Space Sciences,
Shandong University, Weihai 264209, China
}

\begin{abstract}
The recent progress of multiple chiral doublet bands (M$\chi$D) is reviewed for both experimental and theoretical sides. In particular, the experimental findings, theoretical predictions, selection rule for electromagnetic transitions, M$\chi$D with octupole correlations and some related topics are highlighted. Based on these discussions, it is of highly scientific interest to search for the more M$\chi$D as well as the possible chiral wobblers, chirality-parity quartet and chirality-pseudospin triplet (or quartet) bands in nuclear system.
\end{abstract}


\begin{pacs}
      {21.10.Re},              \and
      {21.60.-n},             \and
      {21.60.Ev},

\end{pacs}

\footnotetext[0]{\hspace*{-3mm}\raisebox{0.3ex}{$\scriptstyle\copyright$}2013
Chinese Physical Society and the Institute of High Energy Physics
of the Chinese Academy of Sciences and the Institute
of Modern Physics of the Chinese Academy of Sciences and IOP Publishing Ltd}%

\begin{multicols}{2}

\section{Introduction}\label{section1}

Handedness or chirality is a well-known phenomenon in chemistry, biology and particle physics. Many biological molecules occur in identical but left- and right-handed modes. In particle physics chirality is a dynamical feature of massless particles, which distinguishes the parallel or antiparallel orientation of spin and momentum. However, scientists have long believed that the atomic nucleus are too symmetrical to exist as left- and right-handed versions. The question about the existence of chiral nuclei is thus of great interest.

The first prediction of chirality in atomic nuclei was made by Frauendorf and Meng in 1997~\cite{Frauendorf97}. They pointed out the existence of this phenomenon in triaxial odd-odd nuclei where three angular momentum vectors may couple to each other in either a left- or right-handed mode. Such a chiral geometry may
give rise to pairs of nearly degenerate $\Delta I = 1$ bands with the same parity, i.e., the chiral doublet bands. To test the theoretical prediction of chirality in atomic nuclei, much effort has been devoted to further exploring this interesting phenomenon.
So far, such chiral doublet bands have been reported in the $A \sim$ 80~\cite{Wang11,Liu16,Liu19}, 100~\cite{Vaman04,Joshi04,Timar04,Alcantara04,Timar06,Joshi07,Suzuki08,Kuti14}, 130~\cite{Starosta01,Koike01,Bark01,Hecht01,Hartley01,Mergel02,Koike03,Zhu03,Wang06,Grodner06,Tonev06,Mukhopadhyay07,Zhao09,Grodner11,Ma12,Ayangeakaa13,Ma18}, and 190~\cite{Balabanski04,Lawrie08,Lawrie10,Masiteng13,Masiteng14,Masiteng16,Ndayishimye17} mass regions of the nuclear chart. For details, see recent reviews on nuclear chirality and related topics~\cite{Frauendorf01,Meng08,Meng10,Bark14,Meng16,Raduta16,Starosta17,Frauendorf18,ChenMeng20} or data tables~\cite{Xiong18}.

In 2006, based on the adiabatic and configuration-fixed constrained triaxial relativistic mean field (RMF) calculations, triaxial shape coexistence with high-$j$ proton-hole and neutron-particle configurations is found in $^{106}$Rh, which demonstrates the possible existence of multiple chiral doublet (acronym M$\chi$D) bands in this nucleus~\cite{Meng06}. In the last decade, the theoretical prediction of M$\chi$D has stimulated a lot of experimental efforts~\cite{Ayangeakaa13,Tonev14,Kuti14,Lieder14,Rather14,Liu16,Petrache18,Roy18}. In this review, we will present the recent progress of M$\chi$D.


\section{Experimental findings of M$\chi$D}\label{sec:2.1}

In this section, our attempt is to provide an overview of experimental findings of M$\chi$D.
In 2013, two distinct sets of chiral doublet bands based on the $\pi h^2_{11/2} \otimes \nu h_{11/2}^{-1}$ and $\pi h_{11/2} (g_{7/2})^{-1} \otimes \nu h_{11/2}^{-1}$ configurations were identified in the odd-$A$ nucleus $^{133}$Ce, which was regarded as the strong experimental evidence for the existence of M$\chi$D~\cite{Ayangeakaa13}. The experimental observations of M$\chi$D represent an important confirmation of triaxial shape coexistence and its geometrical interpretation. Later, a novel type of M$\chi$D bands with the identical configuration was found in $^{103}$Rh~\cite{Kuti14}, which shows
that chiral geometry can be robust against the increase of the intrinsic excitation energy.
The M$\chi$D with octupole correlations was identified in the odd-odd $^{78}$Br, which provides the first example of chiral geometry in octupole soft nuclei, and indicates that nuclear chirality can be robust against the octupole correlations~\cite{Liu16}.
It also indicates that a simultaneous breaking of chiral and space-reflection symmetries may exist in nuclei.
In 2018, the M$\chi$D involving 3 and 5 quasiparticle configurations has been observed in odd-A $^{195}$Tl~\cite{Roy18}, which is the first observation of such bands in $A \sim$ 190 mass region. Five pairs of nearly degenerate doublet bands were reported in the even-even $^{136}$Nd~\cite{Petrache18}. Very recently, a new pair of chiral doublet bands with the $\pi g_{7/2}h_{11/2}\otimes \nu h_{11/2}$ configuration was identified in $^{135}$Nd~\cite{Lv19}, which is the isotone of $^{133}$Ce. The new observed chiral doublet bands together with the previously known chiral bands with the $\pi h_{11/2}^{2}\otimes \nu h_{11/2}$ configuration~\cite{Zhu03} constituted a new example of M$\chi$D bands.

\end{multicols}
\begin{table}[ht!]
\footnotesize
\centering
\caption{The observed M$\chi$D nucleus candidate, reaction used to produce M$\chi$D, number of pairs for the observed chiral doublet bands, and single-particle configuration of M$\chi$D}\label{tab1}
\doublerulesep 0.1pt \tabcolsep 13pt 
\begin{tabular}{cccc}
\toprule
  Nucleus  &  Reaction  & Number  & Single-particle configuration \\\hline
  $^{78}$Br~\cite{Liu16}        &  $^{70}$Zn($^{12}$C, p3n)    & 2  & $\pi g_{9/2}\otimes \nu g_{9/2}$, $\pi f_{5/2}\otimes \nu g_{9/2}$\\
  $^{103}$Rh~\cite{Kuti14}      &  $^{96}$Zr($^{11}$B, 4n)     & 3  & $\pi g_{9/2}\otimes \nu h_{11/2}^{2}$, $\pi g_{9/2}\otimes \nu h_{11/2}g_{7/2}$ \\
  $^{105}$Rh~\cite{Alcantara04,Timar04} & $^{100}$Mo($^{11}$B, $\alpha$2n), $^{96}$Zr($^{13}$C, p3n)  & 2  & $\pi g_{9/2}\otimes \nu h_{11/2}^{2}$, $\pi g_{9/2}\otimes \nu h_{11/2}g_{7/2}$ \\
  $^{107}$Ag~\cite{Espinoza97,Jerrestam94,Zhang11} & $^{100}$Mo($^{11}$B, 4n), $^{94}$Zr($^{17}$O, p3n)  & 2  & $\pi g_{9/2}^{-1}\otimes \nu h^2_{11/2}$, $\pi g_{9/2}^{-1}\otimes \nu h_{11/2} (d_{5/2}/g_{7/2})$ \\
  $^{133}$Ce~\cite{Ayangeakaa13}   &  $^{116}$Cd($^{22}$Ne, 5n)$^{133}$Ce  & 2  & $\pi h_{11/2}^{2}\otimes \nu h_{11/2}$, $\pi g_{7/2}h_{11/2}\otimes \nu h_{11/2}$ \\
  $^{135}$Nd~\cite{Zhu03, Lv19}    & $^{110}$Pd($^{30}$Si, 5n), $^{100}$Mo($^{40}$Ar, 5n)  & 2  & $\pi h_{11/2}^{2}\otimes \nu h_{11/2}$, $\pi g_{7/2}h_{11/2}\otimes \nu h_{11/2}$ \\
  $^{136}$Nd~\cite{Petrache18}     & $^{100}$Mo($^{40}$Ar, 5n) & 5  & $\pi h_{11/2}^{1}(d_{5/2},g_{7/2})^{-1}\otimes \nu h_{11/2}^{-1}(s_{1/2},d_{3/2})^{-1}$,\\
                                   &                           &    & $\pi h_{11/2}^{3}(d_{5/2},g_{7/2})^{-1}\otimes \nu h_{11/2}^{-1}(s_{1/2},d_{3/2})^{-1}$,\\
                                   &                           &    & $\pi h_{11/2}^{2}(d_{5/2},g_{7/2})^{-2}\otimes \nu h_{11/2}^{-1}(f_{7/2},h_{9/2})^{1}$,\\
                                   &                           &    & $\pi h_{11/2}^{2}(d_{5/2},g_{7/2})^{-2}\otimes \nu h_{11/2}^{-1}(s_{1/2},d_{3/2})^{-1}$ \\
                                   &                           &    & $\pi h_{11/2}^{2}\otimes \nu h_{11/2}^{-1}(s_{1/2},d_{3/2})^{-1}$ \\
  $^{195}$Tl~\cite{Roy18}          &  $^{185,187}$Re($^{13}$C, xn) & 2  & $\pi h_{9/2}\otimes \nu i_{13/2}^{-2}$, $\pi i_{13/2}\otimes \nu i_{13/2}^{-3}(p_{3/2}f_{5/2})^{-1}$\\

\bottomrule
\end{tabular}
\end{table}
\begin{multicols}{2}

In fact, the existence of more than one chiral configuration in one nucleus had been noticed in 2004. the candidate chiral doublet bands in $^{105}$Rh with $\pi g_{9/2}^{-1}\otimes \nu h_{11/2}^2$ configuration~\cite{Timar04}, and another ones with tentatively suggested $\pi g_{9/2}^{-1}\otimes \nu h_{11/2} (d_{5/2}/g_{7/2})$ configuration~\cite{Alcantara04} were respectively reported by the two different groups. The triaxial RMF approaches have been applied to investigate their triaxial deformations with the corresponding configurations in $^{105}$Rh. Two pairs of doublet bands in $^{105}$Rh were suggested as the candidate M$\chi$D bands~\cite{Li11}. A similar discussion was also applied to $^{107}$Ag~\cite{Qi13}. Two pairs of doublet bands with the $\pi g_{9/2}^{-1}\otimes \nu h^2_{11/2}$ and $\pi g_{9/2}^{-1}\otimes \nu h_{11/2} (d_{5/2}/g_{7/2})$ configurations in $^{107}$Ag have been observed in the experiments~\cite{Espinoza97,Jerrestam94,Zhang11}. The RMF calculations showed the $\pi g_{9/2}^{-1}\otimes \nu h^2_{11/2}$ and $\pi g_{9/2}^{-1}\otimes \nu h_{11/2} (d_{5/2}/g_{7/2})$ bands in $^{107}$Ag have obvious triaxial deformation, $\gamma$ = 27.2$^\circ$ and $\gamma$ = 28.1$^\circ$, respectively. These are favorable deformation parameters for chirality. Using these deformation parameters as input, the multiparticle plus rotor model (MPRM) calculations well reproduced the available data for the two pairs of doublet bands. The chiral geometry of the aplanar rotation for two pairs of doublet bands was further confirmed by analyzing the angular momentum components~\cite{Qi13}.

The observed M$\chi$D nucleus candidate, reaction used to produce M$\chi$D, number of pairs for the observed chiral doublet bands in each nucleus, and single-particle configuration of M$\chi$D are summarized in Tab.~\ref{tab1}.
As shown in Tab.~\ref{tab1}, all M$\chi$D nucleus candidates were discovered in the fusion evaporation reactions using in-beam $\gamma$-ray spectroscopy.
For the single-particle configurations of M$\chi$D, the high-$j$ intruder orbitals (for instance, $g_{9/2}$, $h_{11/2}$ and $i_{13/2}$) are involved in all observed M$\chi$D nucleus candidates. Some low-$j$ orbitals also appeared in the multiparticle configurations, and usually acted as a spectator in the formation of chiral geometry. Furthermore, the observed M$\chi$D nucleus candidates can be roughly divided into two categories. One is M$\chi$D bands with the distinct configurations that differ from each other in their triaxial deformations and configurations. For example, two distinct chiral doublet bands based on the configurations $\pi h_{11/2}^{2}\otimes \nu h_{11/2}$ and $\pi g_{7/2}h_{11/2}\otimes \nu h_{11/2}$ in $^{133}$Ce~\cite{Ayangeakaa13}. The second is M$\chi$D bands with the identical configuration. A unique example is M$\chi$D bands with the $\pi g_{9/2}\otimes \nu h_{11/2}g_{7/2}$ configuration in $^{103}$Rh~\cite{Kuti14}.

\section{Theoretical predictions of M$\chi$D}\label{sec:3}

Theory-wise, M$\chi$D has been investigated with the triaxial PRM~\cite{Droste09,Chen10,Hamamoto13,Zhang16,Jia16}, the combination of triaxial PRM and RMF approaches~\cite{Ayangeakaa13,Kuti14,Liu16,Petrache18,Chen18,Lv19}, the tilted axis cranking model (TAC) with the collective Hamiltonian~\cite{Chen16h,Wu18} and the projected shell model~\cite{Wang19}, etc. In this section I shall only focus on the theoretical predictions of M$\chi$D.

\end{multicols}

\begin{table}[ht!]
\footnotesize
\caption{The calculated triaxial deformation parameters $\beta$ and $\gamma$, corresponding valence nucleon and unpaired nucleon configurations of minima, as well as excitation energies $E_{x}$ in the predicted M$\chi$D nuclei.}
\label{tab2}
\tabcolsep 5pt 
\begin{tabular*}{\textwidth}{cccccccccccccc}
\toprule
 &    &\multicolumn{3}{c}{Configuration} & &($\beta, \gamma$) & $E_{x}$  \\
 \cline{3-5}
Nuclei & & Valence nucleons & & Unpaired nucleons & &   &  (MeV) \\
\hline
$^{54}$Co~\cite{Peng18}& & $\pi f^{-1}_{7/2}\otimes\nu(g^{1}_{9/2}f^{-2}_{7/2})$ & & $\pi f^{-1}_{7/2}\otimes\nu g^{1}_{9/2} $ & & (0.26,18.2$^{\circ}$)& 8.39\\
 & & $\pi (g^{1}_{9/2}f^{-2}_{7/2})\otimes\nu f^{-1}_{7/2}$ & & $\pi g^{1}_{9/2}\otimes\nu f^{-1}_{7/2} $ &&(0.26,17.3$^{\circ}$)& 8.1 \\
\hline
$^{57}$Co~\cite{Peng18}& & $\pi f^{-1}_{7/2}\otimes\nu g^{1}_{9/2}(fp)^{1}$ & & $\pi f^{-1}_{7/2}\otimes\nu g^{1}_{9/2}(fp)^{1} $ &&(0.20,24.0$^{\circ}$)& 5.79\\
& & $\pi f^{-1}_{7/2}\otimes\nu(g^{2}_{9/2})$ & & $\pi f^{-1}_{7/2}\otimes\nu g^{2}_{9/2}$ &&(0.25,36.0$^{\circ}$)&11.82 \\
\hline
$^{60}$Co~\cite{Peng18}& & $\pi f^{-1}_{7/2}\otimes\nu g^{1}_{9/2}(fp)^{4}$ & & $\pi f^{-1}_{7/2}\otimes\nu g^{1}_{9/2}$ & &(0.28,27.0$^{\circ}$)&2.07 \\
 &&$\pi f^{-1}_{7/2}\otimes\nu g^{2}_{9/2}(fp)^{3}$ & & $\pi f^{-1}_{7/2}\otimes\nu g^{2}_{9/2}(fp)^{1}$ &&(0.30,15.1$^{\circ}$)& 6.75\\
\hline
$^{74}$Br~\cite{Qi19}& & $\pi(g^{3}_{9/2}f^{-2}_{5/2}p^{-4}_{3/2}p^{-2}_{1/2})\otimes\nu(g^{4}_{9/2}f^{-1}_{5/2}p^{-2}_{3/2}p^{-2}_{1/2})$ & & $\pi g^{1}_{9/2}\otimes\nu f^{-1}_{5/2} $ &&(0.43,23.2$^{\circ}$)&0.44 \\
& & $\pi(g^{3}_{9/2}f^{-2}_{5/2}p^{-4}_{3/2}p^{-2}_{1/2})\otimes\nu(g^{5}_{9/2}f^{-2}_{5/2}p^{-2}_{3/2}p^{-2}_{1/2})$ & & $\pi g^{1}_{9/2}\otimes\nu g^{-1}_{9/2} $ &&(0.45,27.5$^{\circ}$)& 0.47\\
\hline
$^{76}$Br~\cite{Qi19} & &  $\pi(g^{3}_{9/2}f^{-4}_{5/2}p^{-2}_{3/2}p^{-2}_{1/2})\otimes\nu(g^{5}_{9/2}p^{-2}_{3/2}p^{-2}_{1/2})$ & &  $\pi g^{1}_{9/2}\otimes\nu g^{-1}_{9/2} $ && (0.41,20.8$^{\circ}$)& 0.08\\
& & $\pi(g^{1}_{9/2}f^{-2}_{5/2}p^{-4}_{3/2})\otimes\nu(g^{4}_{9/2}p^{-1}_{3/2}p^{-2}_{1/2})$ & & $\pi g^{1}_{9/2}\otimes\nu p^{-1}_{3/2} $ &&(0.36,32.0$^{\circ}$)& 0.42 \\
& & $\pi(f^{-1}_{5/2}p^{-2}_{3/2}p^{-2}_{1/2})\otimes\nu(g^{5}_{9/2}p^{-2}_{3/2}p^{-2}_{1/2})$ & &$\pi f^{-1}_{5/2}\otimes\nu g^{1}_{9/2} $ &&(0.28,40.3$^{\circ}$)& 2.52 \\
\hline
$^{80}$Br~\cite{Qi19}&&  $\pi(g^{2}_{9/2}f^{-1}_{5/2}p^{-4}_{3/2}p^{-2}_{1/2})\otimes\nu (g^{7}_{9/2}p^{-2}_{1/2})$ & & $\pi f_{5/2}\otimes\nu g^{-1}_{9/2}$ & &   (0.31,23.7$^{\circ}$)& 0.96 \\
& & $\pi(g^{3}_{9/2}f^{-2}_{5/2}p^{-4}_{3/2}p^{-2}_{1/2})\otimes\nu (g^{7}_{9/2}p^{-2}_{3/2})$ & & $\pi g^{1}_{9/2}\otimes\nu g^{-1
}_{9/2}$$\ast$& &  (0.34,25.2$^{\circ}$)& 1.54 \\
\hline
$^{82}$Br~\cite{Qi19}& & $\pi(g^{3}_{9/2}f^{-2}_{5/2}p^{-4}_{3/2}p^{-2}_{1/2})\otimes\nu (g^{7}_{9/2}p^{2}_{3/2}p^{-2}_{3/2})$ & & $\pi g^{1}_{9/2}\otimes\nu g^{-1}_{9/2}$ & &  (0.41,17.5$^{\circ}$)&  6.62 \\
& & $\pi(g^{1}_{9/2}p^{-4}_{3/2}p^{-2}_{1/2})\otimes\nu (g^{7}_{9/2})$ & & $\pi g^{1}_{9/2}\otimes\nu g^{-1}_{9/2}$ & & (0.15,33.6$^{\circ}$)&2.77 \\
& & $\pi(g^{3}_{9/2}f^{-2}_{5/2}p^{-4}_{3/2}p^{-2}_{1/2})\otimes\nu (g^{7}_{9/2})$ & & $\pi g^{1}_{9/2}\otimes\nu g^{-1}_{9/2}$$\ast$
  &&(0.27,10.2$^{\circ}$)&  3.78 \\
\hline
$^{78}$Rb~\cite{Qi18} & & $\pi(g^{3}_{9/2}f^{-4}_{5/2}p^{-2}_{3/2})\otimes\nu(g^{5}_{9/2}f^{-2}_{5/2}p^{-2}_{3/2})$ & &  $\pi g^{1}_{9/2}\otimes\nu g^{-1}_{9/2}$ &  & (0.37,34.0$^{\circ}$)&0.26 \\
& & $\pi(g^{2}_{9/2}f^{-3}_{5/2}p^{-2}_{3/2})\otimes\nu(g^{5}_{9/2}f^{-2}_{5/2}p^{-2}_{3/2})$ & & $\pi f^{1}_{5/2}\otimes\nu g^{-1}_{9/2}$ &&(0.32,38.8$^{\circ}$)&0.51 \\
 & & $\pi(g^{3}_{9/2}f^{-4}_{5/2}p^{-2}_{3/2})\otimes\nu(g^{7}_{9/2}f^{-4}_{5/2}p^{-2}_{3/2})$ & &  $\pi g^{1}_{9/2}\otimes\nu g^{-1}_{9/2}$ &&  (0.40,44.5$^{\circ}$)&1.46\\
\hline
$^{80}$Rb~\cite{Qi18}& & $\pi(g^{1}_{9/2}p^{-2}_{3/2}p^{-2}_{1/2})\otimes\nu(g^{5}_{9/2}p^{-2}_{1/2})$  & &  $\pi g^{1}_{9/2}\otimes\nu g^{-1}_{9/2}$  && (0.22,45.1$^{\circ}$)& 0.31\\
& & $\pi(g^{3}_{9/2}f^{-2}_{5/2}p^{-4}_{3/2})\otimes\nu(g^{6}_{9/2}p^{-1}_{3/2}p^{-2}_{1/2})$ & &$\pi g^{1}_{9/2}\otimes\nu p^{-1}_{3/2}$ &&(0.33,37.3$^{\circ}$)&1.19 \\
& &$\pi(g^{3}_{9/2}p^{-4}_{3/2}p^{-2}_{1/2})$$\otimes$ $\nu(g^{7}_{9/2}p^{-2}_{3/2}p^{-2}_{1/2})$ & &$\pi g^{1}_{9/2}\otimes\nu g^{-1}_{9/2}$ & &(0.35,39.1$^{\circ}$)& 1.98\\
\hline
$^{82}$Rb~\cite{Qi18}& & $\pi(g^{2}_{9/2}p^{-3}_{3/2}p^{-2}_{1/2})\otimes\nu (g^{7}_{9/2}p^{-2}_{1/2}$ )& &$\pi p^{1}_{3/2}\otimes\nu g^{-1}_{9/2}$ & & (0.26,42.7$^{\circ}$)&  0.07\\
& & $\pi(g^{1}_{9/2}p^{-2}_{3/2}p^{-2}_{1/2})\otimes\nu( g^{7}_{9/2}p^{-2}_{1/2}$) & & $\pi g^{1}_{9/2}\otimes\nu g^{-1}_{9/2}$ & & (0.22,36.9$^{\circ}$)&0.24 \\
& & $\pi(g^{3}_{9/2}f^{-2}_{5/2}p^{-4}_{3/2})\otimes\nu (g^{7}_{9/2}p^{-2}_{1/2})$ & &$\pi g^{1}_{9/2}\otimes\nu g^{-1}_{9/2}$ &&(0.29,45.5$^{\circ}$)&0.56
\\
 \hline
 $^{106}$Rh~\cite{Meng06,Peng08,Yao09,Zhao17}& & $\pi g^{-3}_{9/2}\otimes\nu h^{1}_{11/2}(d^{2}_{5/2} or d^{2}_{3/2})$ & & $\pi g^{-1}_{9/2}\otimes\nu h^{1}_{11/2}$$\ast$ & & (0.25,23.3$^{\circ}$)& 0.636\\
& & $\pi g^{-3}_{9/2}\otimes\nu h^{3}_{11/2}$ & & $\pi g^{-1}_{9/2}\otimes\nu h^{1}_{11/2}$ & & (0.30,22.9$^{\circ}$)&1.219 \\
 \hline
 $^{110}$Rh~\cite{Peng08}& & $\pi g^{-3}_{9/2}\otimes\nu h^{3}_{11/2}$ & & $\pi g^{-1}_{9/2}\otimes\nu h^{1}_{11/2}$ & & (0.26,40.6$^{\circ}$)&  0\\
& & $\pi g^{-3}_{9/2}\otimes\nu h^{5}_{11/2}$ & & $\pi g^{-1}_{9/2}\otimes\nu h^{1}_{11/2}$ &&(0.31,18.7$^{\circ}$)&  0.51\\
 \hline

 $^{105}$Ag~\cite{Jia19}& & $\pi (g^{-1}_{9/2}p^{-2}_{1/2})\otimes\nu (h^{1}_{11/2}d^{2}_{5/2}g^{5}_{7/2})$ & & $\pi g^{-1}_{9/2}\otimes\nu h^{1}_{11/2}g^{-1}_{7/2}$$\ast$ &&(0.23,34.9$^{\circ}$)&2.46\\
& & $ (g^{-1}_{9/2}p^{-2}_{1/2})\otimes\nu (h^{1}_{11/2}d^{1}_{5/2}g^{6}_{7/2}$) & & $\pi g^{-1}_{9/2}\otimes\nu h^{1}_{11/2}d^{1}_{5/2}$ &&(0.23,26.5$^{\circ}$)& 3.10\\
& & $ (g^{-1}_{9/2}p^{-2}_{1/2})\otimes\nu (h^{2}_{11/2}g^{6}_{7/2}$) & & $\pi g^{-1}_{9/2}\otimes\nu h^{2}_{11/2}$ &&(0.25,38.5$^{\circ}$)&3.94 \\
 \hline
$^{125}$Cs~\cite{Li18}& & $\pi(g^{4}_{7/2}h^{1}_{11/2})\otimes\nu(sd)^{5}h^{7}_{11/2}$ & & $\pi h^{1}_{11/2}\otimes\nu h^{-1}_{11/2}(sd)^{1}$ & & (0.25,26.3$^{\circ}$)& 2.75 \\
& & $\pi(g^{4}_{7/2}h^{1}_{11/2})\otimes\nu(sd)^{4}h^{8}_{11/2}$ & & $\pi h^{1}_{11/2}\otimes\nu h^{-2}_{11/2}$ &&(0.26,24.3$^{\circ}$)& 4.78\\
 \hline

 $^{129}$Cs~\cite{Li18}& &$\pi(g^{4}_{7/2}h^{1}_{11/2})\otimes\nu(sd)^{7}h^{9}_{11/2}$ & & $\pi h^{1}_{11/2}\otimes\nu h^{-1}_{11/2}(sd)^{1}$ &&(0.21,13.4$^{\circ}$)&2.78 \\
& & $\pi(g^{4}_{7/2}h^{1}_{11/2})\otimes\nu(sd)^{8}h^{8}_{11/2}$ & & $\pi h^{1}_{11/2}\otimes\nu h^{-2}_{11/2}$ &&(0.21,22.7$^{\circ}$)&2.31 \\
 \hline

$^{131}$Cs~\cite{Li18}& & $\pi(g^{4}_{7/2}h^{1}_{11/2})\otimes\nu(sd)^{9}h^{9}_{11/2}$ & & $\pi h^{1}_{11/2}\otimes\nu h^{-1}_{11/2}(sd)^{1}$ & & (0.18,22.0$^{\circ}$)&  2.11\\
& & $\pi(g^{4}_{7/2}h^{1}_{11/2})\otimes\nu(sd)^{10}h^{8}_{11/2}$ & & $\pi h^{1}_{11/2}\otimes\nu h^{-2}_{11/2}$ &&(0.17,24.8$^{\circ}$)&3.00 \\
 \hline

\bottomrule
\end{tabular*}
$\ast$: the configurations have been experimentally observed.
\end{table}

\begin{multicols}{2}

The adiabatic and configuration-fixed constrained triaxial RMF approaches were developed for the first time to investigate the triaxial shape coexistence and possible chiral doublet bands in 2006~\cite{Meng06}. The existence of multiple chiral doublets (M$\chi$D) was suggested in $^{106}$Rh from the examination of the deformation and the corresponding configurations. Similar investigations have also been performed for several isotope chains. These calculations predicted that the M$\chi$D phenomenon might exist in $^{54,57,60}$Co~\cite{Peng18}, $^{74,76,80,82}$Br~\cite{Qi19}, $^{78,80,82}$Rb~\cite{Qi18}, $^{106,110}$Rh~\cite{Peng08,Yao09,Zhao17}, $^{105}$Ag~\cite{Jia19} and $^{125,129,131}$Cs~\cite{Li18} based on the triaxial deformations of the local minima and the corresponding high-$j$ particle(s) and hole(s) configurations.
The predicted multi-chiral nuclei are listed in Tab.~\ref{tab2}, together with the calculated triaxial deformation parameters $\beta$ and $\gamma$, corresponding valence nucleon and unpaired nucleon configurations of minima, as well as excitation energies. Thereinto, the configurations of experimentally observed chiral doublet bands are marked with an asterisk. From Tab.~\ref{tab2}, the excitation energies of most chiral configurations are less than 3 MeV. It is easy to be populated in experiment. The further experimental explorations are highly expected to search for the M$\chi$D in these nuclei.

It is worthwhile to mention that a three-dimensional tilted axis cranking (3DTAC) method based on covariant density functional theory has been recently established and used to investigate the M$\chi$D for the first time in a fully self-consistent and microscopic way~\cite{Zhao17}. This model reproduced well the available experimental spectra and $B(M1)/B(E2)$ ratios in $^{106}$Rh, which exhibited a high predictive power.

\end{multicols}
\begin{center}

\begin{figure}[ht!]
\centering
\includegraphics[bb=80 30 380 180, width=16cm]{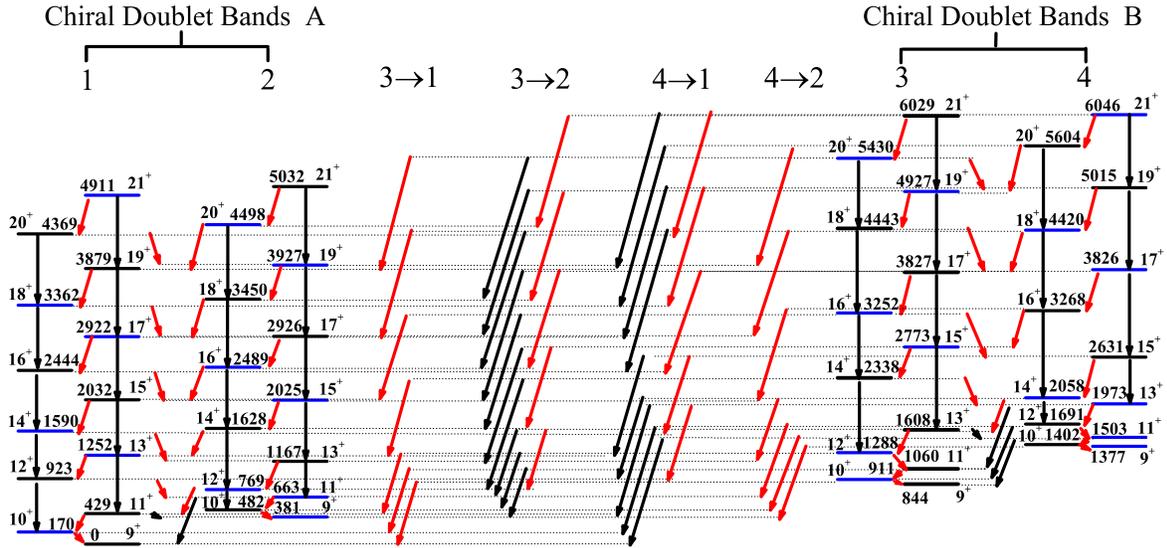}
\caption{Calculated level scheme for M$\chi$D based on the configuration $\pi h_{11/2}\otimes \nu h_{11/2}^{-1}$ coupled with  $\gamma =90^{\circ}$ rotor from Ref.~\cite{Jia16}. Red and black arrows represent $M1(I\rightarrow I-1)$ and $E2(I\rightarrow I-2)$ transitions, respectively.}
\label{fig:1}
\end{figure}

\end{center}

\begin{multicols}{2}

\section{Selection rule for electromagnetic transitions in M$\chi$D}\label{sec:2.3}

Besides the experimental explorations and theoretical predictions of M$\chi$D, it is also interesting to study the fingerprints of M$\chi$D. For a single chiral doublet bands, the known fingerprints for the ideal chirality are as follows: (i) the nearly degeneracy of doublet bands, (ii) the spin independence of $S(I)$, (iii) the similar spin alignments, (iv) the similar $B(M1)$ and $B(E2)$ values, (v) the staggering of $B(M1)$, (vi) the vanishing of the interband $E2$ transitions at high spins, and (vii) the small interaction strength~\cite{Frauendorf97,Koik03,Koik04,Wang07CPL,Liu13}. It is obvious that, for the M$\chi$D with the distinct configurations, the chiral fingerprints are still effective for every pair of chiral doublet bands.

As mentioned above, in comparison with the M$\chi$D that differ from each other in their triaxial deformations and configurations, M$\chi$D may also exist in a single nucleus with the identical configuration.
The chiral bands, including the yrast and yrare bands as well as the higher excited bands, have been studied by the triaxial PRM~\cite{Droste09,Chen10,Hamamoto13,Zhang16}, which has been extensively used in studies of chiral doublet bands and yielded lots of successes~\cite{Peng03,Koik04,Zhang07,Wang07,Wang08,Qi09,Qi091,Wang10,Lawr10,Qi11,Shir12,Chen18}. The PRM calculations~\cite{Droste09,Chen10,Hamamoto13,Zhang16} showed that the properties of the two higher excited bands, including the excitation energies and selection rule for electromagnetic transitions, were very similar to those of the yrast and yrare bands, which indicated that excited doublet bands could be a pair of chiral partners as well.
However, we noted the existence of a number of linking transitions between the lowest-lying chiral bands and higher excited bands. It is necessary to study the properties of these linking transitions.

In Ref.~\cite{Jia16}, the selection rule of electromagnetic transitions for these linking transitions between the lowest and excited chiral doublet bands in M$\chi$D was also studied based on the triaxial PRM. The calculated level scheme for two pairs of chiral doublet bands based on the configuration $\pi h_{11/2}\otimes \nu h_{11/2}^{-1}$ coupled with $\gamma$ =90$^\circ$ rotor was shown in Fig.~\ref{fig:1}.
These bands are labeled as 1,~2,~3 and 4. Bands 1 $\&$ 2 and bands 3 $\&$ 4 form the chiral doublet bands A and B, respectively.
In the calculations, the deformation parameters $\beta = 0.22$, $\gamma = 90^{\circ}$, moment of inertia ${\cal J}_{0} = 30MeV^{-1}\hbar^{2}$, intrinsic quadrupole moment $Q_{0} = (3/\sqrt{5\pi})R^{2}_{0}Z\beta$=3.5 and $g_{p}(g_{n})-g_{R} = 0.7(-0.7)$ are adopted.
As shown in Fig.~\ref{fig:1}, for the in-band $M1$ transitions, the same odd-even spin staggering is clearly seen in the four bands, in which transitions from odd spin to even spin states are allowed. For the interband $M1$ transitions, transitions from band 1 (3) to 2 (4) and band 2 (4) to 1 (3) are allowed for even spin states decaying to odd spin states. The linking $M1$ transitions between chiral doublet bands A and B are allowed from the states of the band 3 decaying alternatively to those of the band 1 or 2, then the same behavior for band 4 is exhibited. Based on these calculated results, a whole selection rules of electromagnetic transitions including in-band, interband and linking transitions were summarized in Ref.~\cite{Jia16} (see the Tab.~3 in Ref.~\cite{Jia16}). Besides the whole selection rules, the quantitative relations of electromagnetic transitions probabilities were also obtained in Ref.~\cite{Jia16}. The $B(M1)$ and $B(E2)$ values in the excited chiral doublet bands have the same order of magnitude as those in lowest chiral doublet bands. However, the $B(M1)$ and $B(E2)$ values of transitions which link the excited to the lowest chiral doublet bands are two orders of magnitude smaller than those in the lowest (or excited) chiral doublet bands. The selection rules and the quantitative relations of electromagnetic transitions probabilities would be helpful for confirming the existence of M$\chi$D bands with the identical configuration in the real nuclei.

In Fig.~\ref{fig:1}, the interband $E2$ transitions in chiral doublet bands A (or B) are forbidden, which is consistent with the fingerprint (vi) of ideal chiral doublet bands. However, there exist a number of $E2$ linking transitions between chiral doublet bands A and B. It implies that such bands 2 and 3 can't be defined as a pair of ideal chiral partners. The existence of $E2$ linking transitions allow us to extract the interaction strength between chiral doublet bands A and B. For the doublet bands with the same configuration and deformation, the interaction strength $V$ can be expressed as~\cite{Liu13}
\begin{equation}
V=\frac{\sqrt{R}\Delta E_{I}\Delta E_{I-2}}{\sqrt{(R+1)^{2}(\Delta E_{I}^{2}+\Delta E_{I-2}^{2})+2\Delta E_{I}\Delta E_{I-2}(R^{2}-1)}} ,
\end{equation}
where $R=\frac{B[E2,I^{yrare}\rightarrow (I-2)^{yrast}]}{B[E2,I^{yrare}\rightarrow (I-2)^{yrare}]}$.
According to the Eq.~(1), we extracted the interaction strength between chiral doublet bands A and B. The values of $B(E2)$ and $\Delta E$ come from the calculated results of PRM~\cite{Jia16}. The calculations show that the average interaction strength $\overline{V}$ between chiral doublet bands A and B is approximately equal to 200 keV in the chiral range, which implies that the chiral geometry is mixed by the vibrational component~\cite{Liu13,Qi10}.
It should be noted that the Eq.~(1) are deduced from the two-band mixing picture. Thus, the present calculated V is an approximate solution to M$\chi$D bands.

In order to show the picture more clearly, the mixing ratio and the percentage of $E2$ mixing for the $\Delta I$=1 linking transitions are calculated in the present work by the PRM with the ideal case i.e. the configuration $\pi h_{11/2}\otimes \nu h_{11/2}^{-1}$ with $\gamma$ =90$^\circ$. The calculations show $\sim$1\% and 2\% $E2$ admixture for the $\Delta I$=1 transitions in the chiral doublet bands A and B, respectively, suggesting these transitions as being essentially of a pure $M1$ character. Moreover, relatively large $E2$ fractions ($\sim$20\%) are obtained for the $\Delta I$=1 linking transitions between the chiral doublet bands A and B. The enhance $E2$ component is a characteristic feature of the wobbling phonon excitation~\cite{SW01}. The similar conjecture has been obtained in Ref.~\cite{Koike05} based on the same model calculations. Instead of analyzing the interaction strength between chiral doublet bands A and B, Ref.~\cite{Koike05} studied the expectation value $\sqrt{\langle R _{3}^{2}\rangle}$ of the core rotation along the quantization axis, and found that components of the core rotation along the long
and short axis for the excited doublet bands are larger than those for the lowest chiral bands. It implied that the excited pair exhibited a chiral
geometry which is realized with a wobbling motion of the core. Hence, the excited pair was claimed as the chiral wobblers~\cite{Koike05}.
Further detailed studies beyond the scope of this paper are needed to study whether the excited doublet bands are associated with the chiral wobblers.
The possible coexistence of chirality and the other rotational modes will be discussed in the following section.

\section{M$\chi$D with octupole correlations and some related topics}\label{sec:5}

If another spontaneous breaking of discrete symmetry takes place in addition to chirality, the degree of energy degeneracy will accordingly be increased resulting in multiple degenerate $\Delta I$=1 rotational bands. For instance, in the case of the chiral and reflection symmetry breakings, parity and chiral doubling bring about a set of four degenerate $\Delta I$=1 bands. The first such example, though rather soft breaking of reflection symmetry, has been found in
$^{78}$Br~\cite{Liu16}. In Ref.~\cite{Liu16}, two pairs of positive- and negative-parity doublet bands together with eight strong electric dipole transitions linking their yrast positive- and negative-parity bands in $^{78}$Br have been found. It provided the evidence for M$\chi$D bands with octupole correlations, reported the first example of chiral geometry in octupole soft nuclei, and indicated that nuclear chirality can be robust against the octupole correlations.
This observation also pointed to the exciting possibility of observing the chirality-parity quartet (CPQ) bands i.e., four
$\Delta I$=1 alternating-parity rotational bands with the same configuration in a single nucleus with both stable triaxial and octupole deformations.
So far, CPQ bands have not been experimentally observed.

Note that, a pair of positive-parity doublet bands and several E1 transitions linking yrast positive- and negative-parity bands in $^{124}$Cs have been reported by Ref.~\cite{Gizon01} and Refs.~\cite{Lu00,Dong09}, respectively. Based on the TPRM calculations, Ref.~\cite{Wang10} suggested that the positive-parity doublet bands in $^{124}$Cs might correspond to a typical chiral vibration pattern. Recently, lifetime measurements have been carried out using the Doppler shift attenuation method (DSAM) for the yrast positive- and negative-parity bands in $^{124}$Cs~\cite{Selvakumar15}. The measured results~\cite{Selvakumar15} show that the $B(E1)$ rates are of the order of 10$^{-4}$ W.u., thereby indicating coexistence of a pair of chiral doublet bands and octupole correlations in $^{124}$Cs.

\begin{figure}[ht!]
\centering
\includegraphics[bb=30 240 490 460, width=14cm]{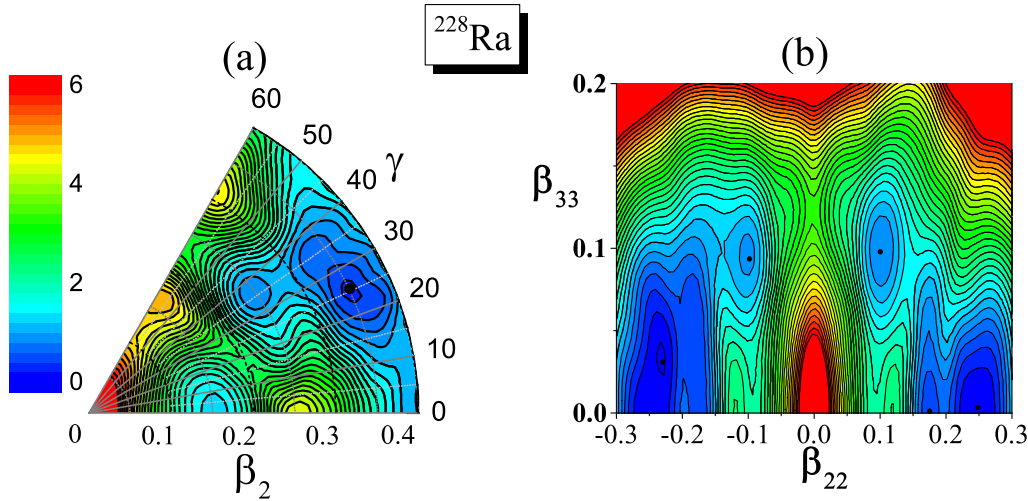}
\caption{The calculated PES of $^{228}$Ra in the $\beta_{2}-\gamma$ (a) and $\beta_{22}-\beta_{33}$ (b) planes. The energies are normalized with respect to the ground state. The contour separation is 0.2 MeV.}
\label{fig:2}
\end{figure}

In order to search for the possibly candidate cores to construct CPQ bands, the potential energy surfaces (PES) of the even-even Se, Ba, and Ra isotopes were calculated by using the macroscopic-microscopic method in a multidimensional space $\{\alpha_{\lambda, \mu}\}$ including quadrupole ($\lambda$=2, $\mu$=0, 2)
and octupole ($\lambda$=3, $\mu$=0, 1, 2, 3) degrees of freedom~\cite{Liu18}.
The calculated results showed that the even-even isotopes $^{92}$Se, $^{112,114,144-150}$Ba and $^{220-228}$Ra can exhibit the coexistence of triaxial and octupole deformations. It is therefore expected that CPQ bands can be observed experimentally in these even-even nuclei and their neighboring odd-A/odd-odd nuclei. As an example, the calculated PES of $^{228}$Ra in the $\beta_{2}-\gamma$ and $\beta_{22}-\beta_{33}$ planes using the macroscopic-microscopic method~\cite{Nazarewicz84,Dudek04,Mazurek05} are shown in Fig.~\ref{fig:2}(a) and \ref{fig:2}(b), respectively. One can see from Fig.~\ref{fig:2} that $^{228}$Ra has the obviously triaxial and octupole deformations.

It is necessary to note here that in Tab.~\ref{tab1}, the configurations of some M$\chi$D involve the orbits of pseudospin doublet states (e.g. 1g$_{7/2}$, 2d$_{5/2}$). Thus, a competing interpretation of these doublet bands would include the pseudospin doublet bands.
Pseudospin symmetry in atomic nuclei was introduced in 1969~\cite{Arima69,Hecht69}. A pair of nearly degenerate doublet bands with the configuration involving pseudospin doublet states have been observed and suggested as the pseudospin doublet bands in several nuclei~\cite{108Tc,108Ag1,108Ag2,118Sb,126Cs,189Pt,116Sb,125Cs,Wang16}.

\end{multicols}

\begin{multicols}{2}

A specific calculation~\cite{Jia19} for the nearly degenerate triplet bands with the $\pi g_{9/2}^{-1}\otimes \nu h_{11/2} (d_{5/2}/g_{7/2})$ configuration
in $^{105}$Ag was performed by using the RMF theory and the MPRM. The configuration-fixed constrained
triaxial RMF calculations exhibited the pseudospin symmetry in single particle spectra and triaxial shape coexistence. The experimental excitation energies and electromagnetic transition probabilities for the triplet bands were well reproduced by the MPRM calculations. Thus, the first $\&$ second lowest energy bands and the second $\&$ third bands were interpreted as the pseudospin doublet bands and chiral doublet bands, respectively. This work also motivated the investigation to search for the chirality-pseudospin triplet (or quartet) bands in the nuclei.

\section{Summary and perspectives}\label{sec:6}
The recent progress in M$\chi$D is reviewed for both experimental and theoretical sides. In particular, the experimental findings, theoretical predictions, selection rule of electromagnetic transitions, M$\chi$D with octupole correlations and some related topics are highlighted.
Based on the above discussion, it is of highly scientific interest to search for the more M$\chi$D as well as the possible chiral wobblers, chirality-parity quartet and chirality-pseudospin triplet (or quartet) bands in nuclear system.
On the other hand, these exotic nuclear phenomena have brought severe challenges to current nuclear models and, thus, require the development of new approaches.
Very recently, to study the M$\chi$D with octupole correlations in $^{78}$Br, a reflection-asymmetric triaxial PRM with a quasi-proton and a quasi-neutron coupled with a reflection-asymmetric triaxial rotor has been developed ~\cite{Wangyy19}.
According to the present review, we would also like to attract more experimental and theoretical efforts on the investigation of chirality or multiple chirality in the atomic nucleus.

\emph{The author is grateful to B. Qi, C. Liu, H. Jia, and N. B. Zhang for helpful discussions and careful readings of the manuscript. }

\end{multicols}
\vspace{-1mm}
\centerline{\rule{80mm}{0.1pt}}
\vspace{2mm}

\begin{multicols}{2}

\end{multicols}

\clearpage
\end{CJK*}
\end{document}